\input{aipcheck}

\documentclass[
    ,final            
  ]
  {aipproc}

\layoutstyle{6x9}

\makeatletter
\def\slash#1{{\mathpalette\c@ncel{#1}}} 
\makeatother

\newcommand\beq{\begin{eqnarray}}
\newcommand\eeq{\end{eqnarray}}

\newcommand\la{\langle}
\newcommand\ra{\rangle}

\begin{document}

\title{Probing multigluon correlations through single
spin asymmetries in pp collisions}

\classification{12.38.Bx, 13.85.Ni, 13.88.+e}
\keywords      {single spin asymmetry, twist-3, three-gluon correlation
function}

\author{Yuji Koike}{
  address={Department of Physics, Niigata University,
Ikarashi, Niigata 950-2181, Japan}
}

\author{Shinsuke Yoshida}{
  address={Graduate School of Science and Technology, Niigata University,
Ikarashi, Niigata 950-2181, Japan}
}

\begin{abstract}
We study the single spin asymmetries
for the $D$-meson production, Drell-Yan lepton-pair
production and the direct-photon production
in the $pp$ collision induced by the twist-3 three-gluon correlation functions
in the transversely polarized nucleon.
We present a corresponding polarized cross section formula in the leading-order with respect to
the QCD coupling 
and a model calculation for the asymmetries, 
illustrating the sensitivity to the form of the correlation functions. 
\end{abstract}

\maketitle


\section{1. Introduction}

Understanding the origin of the large single spin asymmetries (SSAs) observed in various
high-energy semi-inclusive processes have been a big challenge during
the past decades. The SSA can be generated as a consequence of the
multiparton correlations inside the hadrons in the collinear
factorization approach which is valid when the transverse momentum of the
particle in the final state can be regarded as hard. 
Among such multiparton correlations, purely gluonic correlations have a potential importance, 
since gluons are ample in the nucleon.  
The best way to probe such gluonic correlations is the measurement of SSA for a heavy meson
production in semi-inclusive deep inelastic scattering (SIDIS) 
and the $pp$ collision\,\cite{KQVY08,BKTY10}, since the
heavy-quark pair one of which fragments into the final-state meson is mainly produced, respectively, by the
photon-gluon and gluon-gluon fusion mechanisms.
The measurement of SSA for the $D$-meson production in the $pp$ collision is
ongoing at RHIC\,\cite{Liu}.

In this report, we study the contribution of the three-gluon correlation functions
representing such multigluonic effects
to SSA in the $pp$ collision for the $D$-meson production ($p^\uparrow p\to DX$)\,\cite{KY11},
Drell-Yan lepton-pair production ($p^\uparrow p\to \gamma^*X$) and 
the direct-photon production ($p^\uparrow p\to \gamma X$)\,\cite{KY11direct}.  
We will present the corresponding single-spin dependent cross sections 
by applying the formalism 
developed for SIDIS, $ep^\uparrow\to eDX$\,\cite{BKTY10}.
We will also present a model calculation of the 
SSAs for $p^\uparrow p\to DX$ and $p^\uparrow p\to\gamma X$ induced by the
three-gluon correlation functions
in comparison with the RHIC preliminary data for the former\,\cite{Liu}.

\section{2. Three-gluon correlation functions}

Three-gluon correlation functions in the transversely 
polarized nucleon are defined as the color-singlet nucleon matrix element
composed of the three gluon's field strength tensors $F^{\alpha\beta}$.
Corresponding to the two structure constants for the color SU(3) group, 
$d_{bca}$ and $f_{bca}$, 
one obtains two independent three-gluon correlation functions $O(x_1,x_2)$ and $N(x_1,x_2)$ 
as\,\cite{BKTY10}
\small
\beq
&&\hspace{-0.8cm}O^{\alpha\beta\gamma}(x_1,x_2)
=-g(i)^3\int{d\lambda\over 2\pi}\int{d\mu\over 2\pi}e^{i\lambda x_1}
e^{i\mu(x_2-x_1)}\la pS|d_{bca}F_b^{\beta n}(0)F_c^{\gamma n}(\mu n)F_a^{\alpha n}(\lambda n)
|pS\ra \nonumber\\
&&=2iM_N\left[
O(x_1,x_2)g^{\alpha\beta}\epsilon^{\gamma pnS}
+O(x_2,x_2-x_1)g^{\beta\gamma}\epsilon^{\alpha pnS}
+O(x_1,x_1-x_2)g^{\gamma\alpha}\epsilon^{\beta pnS}\right]
\label{3gluonO},\\
&&\hspace{-0.8cm}N^{\alpha\beta\gamma}(x_1,x_2)
=-g(i)^3\int{d\lambda\over 2\pi}\int{d\mu\over 2\pi}e^{i\lambda x_1}
e^{i\mu(x_2-x_1)}\la pS|if_{bca}F_b^{\beta n}(0)F_c^{\gamma n}(\mu n)F_a^{\alpha n}(\lambda n)
|pS\ra \nonumber\\
&&=2iM_N\left[
N(x_1,x_2)g^{\alpha\beta}\epsilon^{\gamma pnS}
-N(x_2,x_2-x_1)g^{\beta\gamma}\epsilon^{\alpha pnS}
-N(x_1,x_1-x_2)g^{\gamma\alpha}\epsilon^{\beta pnS}\right],
\label{3gluonN}
\eeq
\normalsize
where $S$ is the transverse-spin vector 
for the nucleon, 
$n$ is the light-like vector satisfying $p\cdot n=1$, and 
we have used the shorthand notation as $F^{\beta n}\equiv
F^{\beta\rho}n_{\rho}$ {\it etc}.  The gauge-link operators which
restore gauge invariance of the correlation functions
are suppressed in (\ref{3gluonO}) and (\ref{3gluonN}) for simplicity.  
The nucleon mass $M_N$ is introduced to define $O(x_1,x_2)$ and $N(x_1,x_2)$ dimensionless.

\section{3. $D$-meson production in $pp$ collision}

Applying the formalism for
the contribution of the three-gluon correlation functions to SSA developed in \cite{BKTY10}, 
the twist-3 cross section for
$p^{\uparrow}(p,S_\perp) + p(p') \to D(P_h)  + X$
with the center-of-mass energy $\sqrt{S}$ can be obtained in the following form\,\cite{KY11}: 
\footnotesize
\beq
&&\hspace{-0.9cm}
P_h^0\frac{d\sigma^{{\rm tw3},D}}{d^3P_h}=\frac{\alpha_s^2M_N\pi}{S}\epsilon^{P_h p n S_{\perp}}
\sum_{f=c\bar{c}}\int\frac{dx'}{x'}G(x')\int\frac{dz}{z^2}D_f(z)\int\frac{dx}{x}\delta
 \left(\tilde{s}+\tilde{t}+\tilde{u}\right){1\over z\tilde{u}}
 \nonumber\\
&&\hspace{-0.9cm}
\times\biggl[\delta_f\left\{
\left(\frac{d}{dx}O(x,x)-\frac{2O(x,x)}{x}\right)\hat{\sigma}^{O1}
+\left(\frac{d}{dx}O(x,0)-\frac{2O(x,0)}{x}\right)\hat{\sigma}^{O2}
+\frac{O(x,x)}{x}\hat{\sigma}^{O3}
+\frac{O(x,0)}{x}\hat{\sigma}^{O4}
\right\} \nonumber\\
&&\hspace{-0.9cm}
+\left\{
\left(\frac{d}{dx}N(x,x)-\frac{2N(x,x)}{x}\right)\hat{\sigma}^{N1}
+\left(\frac{d}{dx}N(x,0)-\frac{2N(x,0)}{x}\right)\hat{\sigma}^{N2}
+\frac{N(x,x)}{x}\hat{\sigma}^{N3}
+\frac{N(x,0)}{x}\hat{\sigma}^{N4}
\right\}
\biggr],
\label{twist3final}
\eeq
\normalsize
where 
$\delta_c=1$ and 
$\delta_{\bar{c}}=-1$, $D_f(z)$ represents the 
$c\to D$ or $\bar{c}\to\bar{D}$ fragmentation functions, $G(x')$ is the
unpolarized gluon density, $p_c$ is the four-momentum of the $c$ (or
$\bar{c}$) quark (mass $m_c$) fragmenting into the
final $D$ (or $\bar{D}$) meson and 
$\tilde{s}$, $\tilde{t}$, $\tilde{u}$ are defined as $
\tilde{s}=(xp+x'p')^2,~\tilde{t}=(xp-p_c)^2-m_c^2,~\tilde{u}=(x'p'-p_c)^2-m_c^2.$
The hard cross sections $\hat{\sigma}^{O1,O2,O3,O4}$ and
$\hat{\sigma}^{N1,N2,N3,N4}$ are listed in \cite{KY11}.  
The cross section (\ref{twist3final}) receives contributions from
$O(x,x)$, $O(x,0)$, $N(x,x)$ and $N(x,0)$ separately, which differs from the previous
result\,\cite{KQVY08}. 

We perform numerical estimate for $A_N$ based on (\ref{twist3final}).
For the RHIC kinematics, we found that the terms with $\hat{\sigma}^{O3,O4,N3,N4}$
are negligible compared with those with $\hat{\sigma}^{O1,O2,N1,N2}$
and that
$\hat{\sigma}^{O1} \simeq \hat{\sigma}^{O2} \sim \hat{\sigma}^{N1} \simeq -\hat{\sigma}^{N2}$.  
One can thus regard the cross section as a function of the correlation functions
$O(x)\equiv O(x,x)+O(x,0)$ and $N(x)\equiv N(x,x)-N(x,0)$ to a very good approximation.  
We assume the relation 
$O(x)=\pm N(x)$ together with $O(x,x)=O(x,0)$ and $N(x,x)=-N(x,0)$ for simplicity. 
For the functional form of each function, we employ the following two models:  
\beq
&&{\rm Model\ 1}:\qquad O(x)=K_G\,x\,G(x),
\label{model1}\\
&&{\rm Model\ 2}:\qquad O(x)=K_G'\,\sqrt{x}\,G(x), 
\label{model2}
\eeq
where $K_G$ and $K_G'$ are the constants to be determined so that the
calculated asymmetry is consistent with the RHIC data\,\cite{Liu}.

For the numerical calculation, we use GJR08 \cite{GJR08}
for $G(x)$ and KKKS08 \cite{KKKS08} for $D_f(z)$.  We calculate $A_N$
for the $D$ and
$\bar{D}$ mesons at the RHIC energy at $\sqrt{S}=200$
GeV and the transverse momentum of the $D$-meson $P_T=2$ GeV.
We set the scale of all the distribution and fragmentation functions at
$\mu=\sqrt{P_T^2+m_c^2}$ with the charm quark mass $m_c=1.3$ GeV.

\begin{figure}[h]
\scalebox{0.35}{\includegraphics{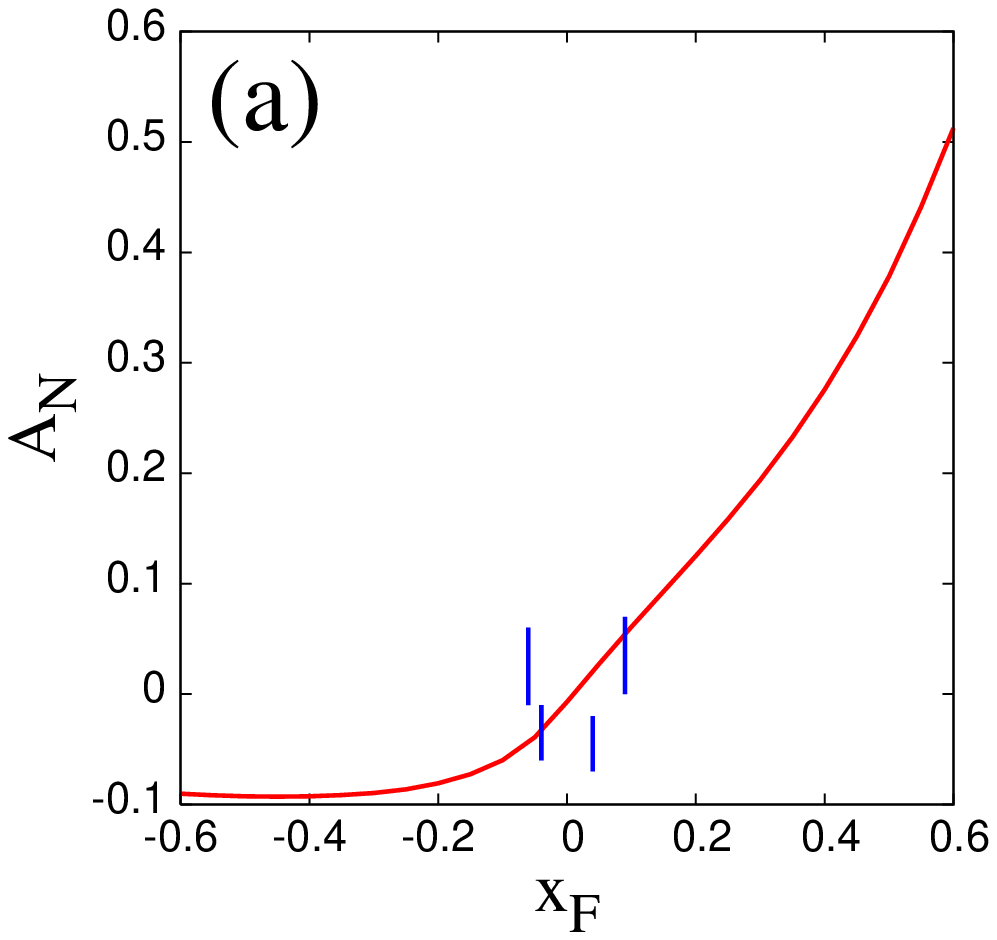}}
\scalebox{0.35}{\includegraphics{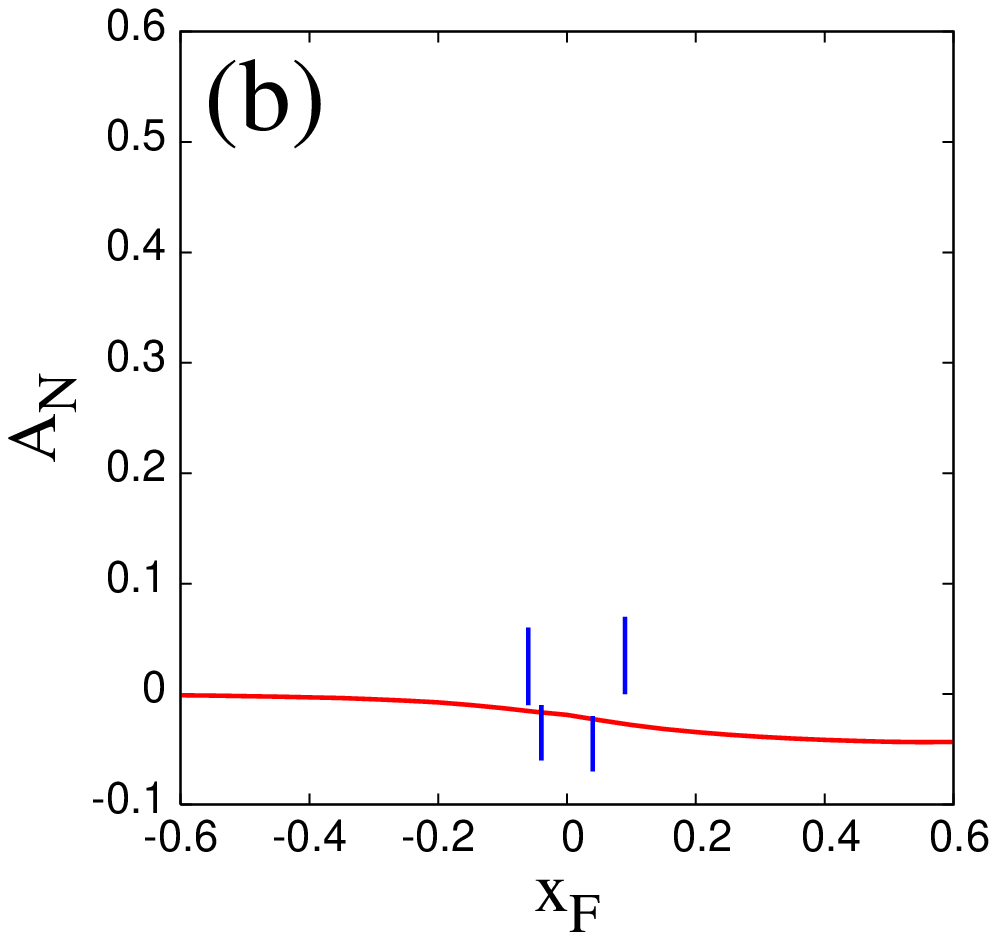}}
\scalebox{0.35}{\includegraphics{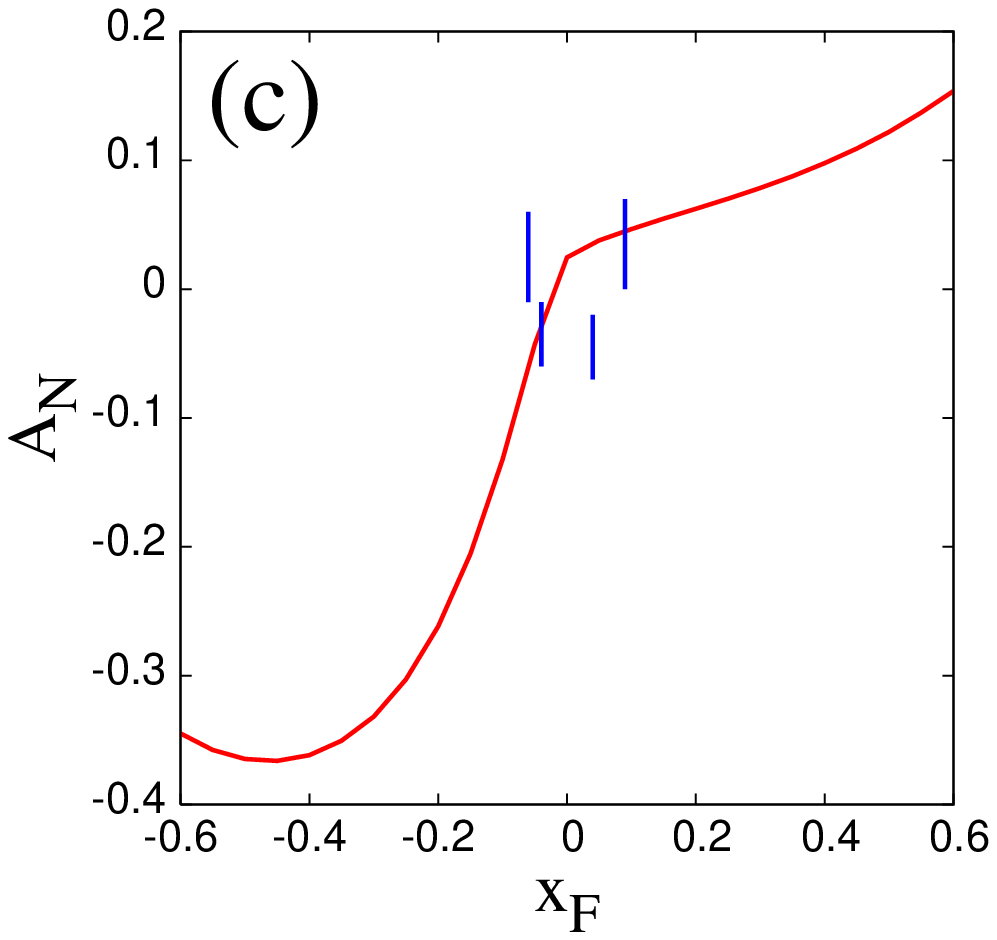}}
\scalebox{0.35}{\includegraphics{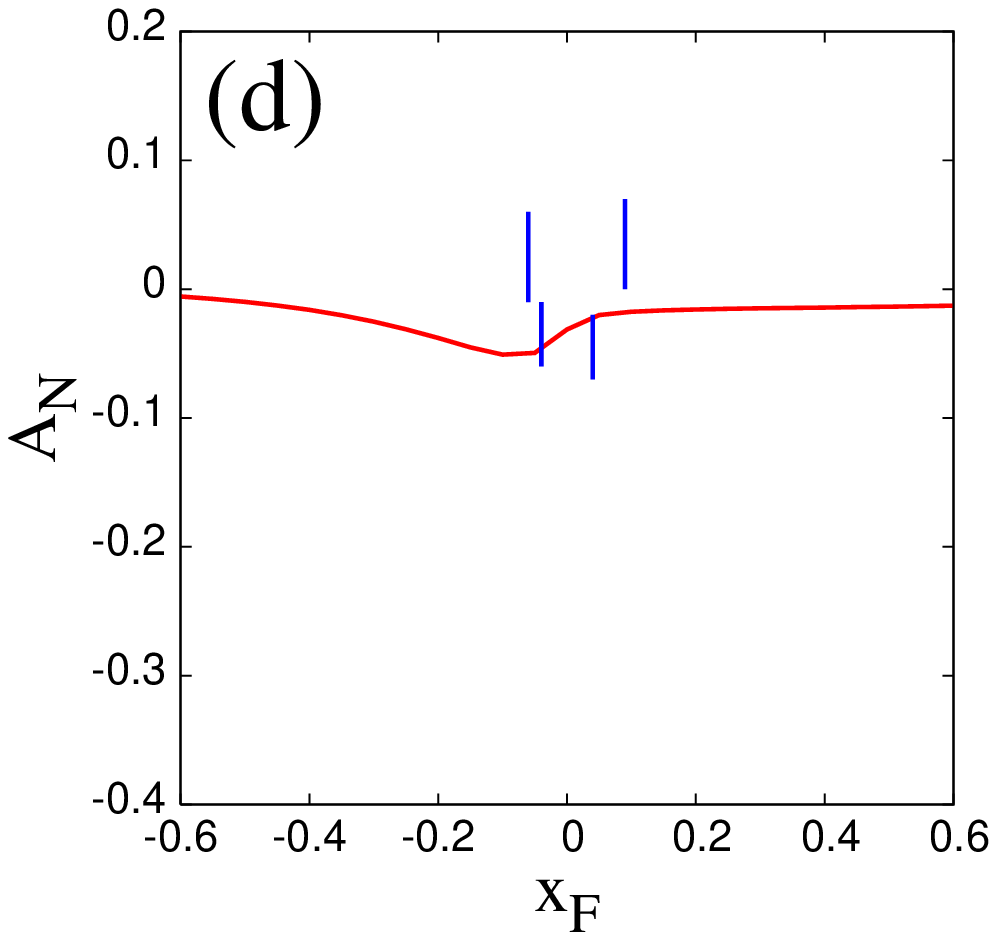}} 
 \caption{(a) $A_N^D$ for $D^0$ and (b) $A_N$ for $\bar{D^0}$ 
 for Model 1 in (\ref{model1}) with $O(x)=N(x)$ and $K_G=0.004$, and (c) $A_N^D$ for
 $D^0$ and (c) $A_N$ for $\bar{D^0}$ for Model 2 in
 (\ref{model2}) with $O(x)=N(x)$ and $K_G'=0.001$.  Short bars denote the RHIC preliminary data
 taken from \cite{Liu}.}
 \label{fig:one}
\end{figure}

Fig. 1 shows the result of $A_N$ for the $D^0$ and $\bar{D}^0$ 
mesons with the relation $O(x)=N(x)$
together with the preliminary data\,\cite{Liu} denoted by the
short bars.  
The sign of the contribution from $O(x)$ changes
between $D^0$ and $\bar{D}^0$ as shown in (\ref{twist3final})
and works constructively (destructively) for $D^0$ ($\bar{D}^0$)
for the case $O(x)=N(x)$.  
The values $K_G=0.004$ and $K'_G=0.001$ have been determined so that
$A_N$ does not overshoot the RHIC data.
If one adopts the relation $O(x)=-N(x)$, 
the result for the $D^0$ and $\bar{D}^0$ mesons will be interchanged.  
The rising behavior of $A_N$ at large $x_F$ as shown in Figs. 1(a) and (c) is
originated from the derivative of $O(x)$ and $N(x)$, as in the case of
the soft-gluon-pole (SGP) contribution for the quark-gluon correlation function. 
By comparing the results for the models 1 and 2 in Figs. 1(a) and (c), 
one sees that the behavior of $A_N$ at $x_F<0$ depends strongly on
the small-$x$ behavior of the three-gluon correlation functions.  
Therefore $A_N$ at $x_F<0$ is useful to get constraint on the
small-$x$ behavior of the three-gluon correlation functions.  

\section{4. Drell-Yan lepton-pair production in $pp$ collision}

The analysis of the previous section can be extended to
$A_N$ for the Drell-Yan lepton-pair production induced by the three-gluon correlation function.
The corresponding twist-3 diagrams for the hard part are shown in Fig. 2,
which give rise to SGP contribution at $x_1=x_2$ due to the initial-state interaction
between the extra incoherent gluon from the polarized proton and the quark coming out of the
unpolarized proton.  
\begin{figure}[h]

\scalebox{0.5}{\includegraphics{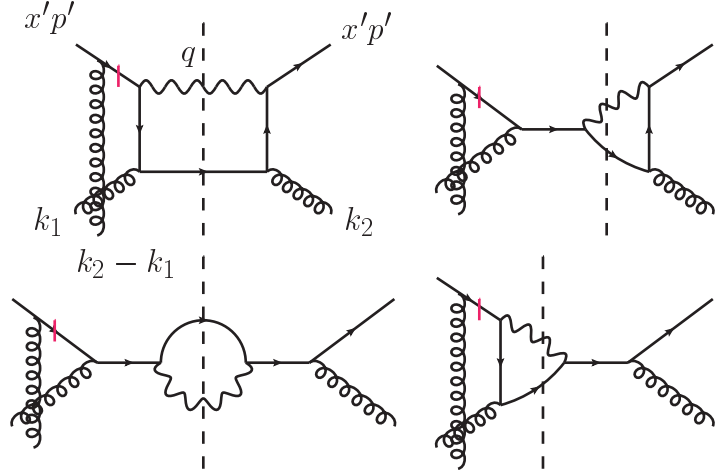}}

\caption{The diagrams for the three-gluon contribution to 
the twist-3 cross section for $p^\uparrow p\to\gamma^* X$.
In the collinear limit, the momenta $k_1$ and $k_2$ 
coming from the polarized nucleon are set to $k_i=x_ip$ ($i=1,2$)
and the pole contribution from the bared propagator gives rise to the SSA.}
\end{figure}
From these diagrams, one obtains for the single-spin-dependent cross section 
for the Drell-Yan process, 
$p^{\uparrow}(p,S_\perp) + p(p') \to \gamma^*(q)  + X$,
with the invariant mass $Q^2=q^2$ for the lepton-pair as\,\cite{KY11direct}
\beq
&&{d\sigma^{\rm tw3,DY}\over dQ^2dyd^2\vec{q}_{\perp}}={2 M_N\alpha_{em}^2\alpha_s\over
3 SQ^2}\int {dx\over x}\int {dx'\over x'}\delta
(\hat{s}+\hat{t}+\hat{u}-Q^2)\epsilon^{q p n S_{\perp}}{1\over \hat{u}}\sum_ae_a^2f_a(x')
 \nonumber\\
&&\qquad\qquad
\times\biggl[\delta_a\left\{
\Bigl(\frac{d}{dx}O(x,x)-\frac{2O(x,x)}{x}\Bigr)\hat{\sigma}_1
+\Bigl(\frac{d}{dx}O(x,0)-\frac{2O(x,0)}{x}\Bigr)\hat{\sigma}_2
\right.\nonumber\\
&&\left.\qquad\qquad\qquad
+{O(x,x)\over
x}\hat{\sigma}_3+{O(x,0)\over x}\hat{\sigma}_4 \right\}\nonumber\\
&&\qquad\qquad
-\Bigl(\frac{d}{dx}N(x,x)-\frac{2N(x,x)}{x}\Bigr)\hat{\sigma}_1
+\Bigl(\frac{d}{dx}N(x,0)-\frac{2N(x,0)}{x}\Bigr)\hat{\sigma}_2
\nonumber\\
&&\qquad\qquad\qquad-{N(x,x)\over
x}\hat{\sigma}_3+{N(x,0)\over x}\hat{\sigma}_4 \biggr],
\label{DY}
\eeq
where $y$ is the rapidity of the virtual photon, $\alpha_{em}\simeq 1/137$
is the QED coupling constant, and the partonic hard cross sections are defined as
\beq
\hat{\sigma}_{1}={2\over N}({\hat{u}\over \hat{s}}+{\hat{s}\over
\hat{u}}+{2Q^2\hat{t}\over \hat{s}\hat{u}}), \qquad
\hat{\sigma}_{2}={2\over N}({\hat{u}\over \hat{s}}+{\hat{s}\over
\hat{u}}+{4Q^2\hat{t}\over \hat{s}\hat{u}}), \nonumber\\
\hat{\sigma}_{3}=-{1\over N}{4Q^2(Q^2+\hat{t})\over \hat{s}\hat{u}},\qquad
\hat{\sigma}_{4}=-{1\over N}{4Q^2(3Q^2+\hat{t})\over \hat{s}\hat{u}},
\eeq
with the number of colors $N=3$ and 
$\hat{s}=(xp+x'p')^2$, $\hat{t}=(xp-q)^2$ and $\hat{u}=(x'p'-q)^2$.  
For a large $Q^2$, $\hat{\sigma}_{1}$ differs from $\hat{\sigma}_{2}$ significantly, and
$\hat{\sigma}_{3,4}$ are not negligible.  Therefore the cross section (\ref{DY}), in general, 
depends on the four functions $O(x,x)$, $O(x,0)$, $N(x,x)$ and $N(x,0)$ independently
unlike the case for $p^\uparrow p\to DX$ 
in the previous section, where the twist-3 cross section can be regarded as a function of
the combination $O(x,x)+O(x,0)$ and $N(x,x)-N(x,0)$. 

The $A_N$ for the Drell-Yan process receives contribution not only from the
three-gluon correlation functions but also from the quark-gluon correlation
functions, for which the twist-3 cross section have been derived in \cite{JQVY,KT07,KK11}. 
The sum of (\ref{DY}) and those from the quark-gluon correlation functions
gives the complete leading-order cross section for the asymmetry.

\section{5. Direct photon production in $pp$ collision}

Taking the $q\to 0$ limit of (\ref{DY}), 
the twist-3 cross section for the direct photon production, 
$p^{\uparrow}(p,S_\perp) + p(p') \to \gamma(q)  + X$, 
induced by the three-gluon correlation functions
can be obtained as\,\cite{KY11direct}
\small
\beq
 E_{\gamma}\frac{d\sigma^{\rm tw3,DP}}{d^3q}&=&\frac{4\alpha_{em}\alpha_sM_N\pi}{S}
\sum_{a}e_a^2 \int\frac{dx'}{x'}f_a(x')\int\frac{dx}{x}\delta
 (\hat{s}+\hat{t}+\hat{u})\epsilon^{q p n S_{\perp}} \nonumber\\
&& \times\biggl[\delta_a
\Bigl(\frac{d}{dx}O(x,x)-\frac{2O(x,x)}{x}
+\frac{d}{dx}O(x,0)-\frac{2O(x,0)}{x}\Bigr) \nonumber\\
&&
-\frac{d}{dx}N(x,x)+\frac{2N(x,x)}{x}
+\frac{d}{dx}N(x,0)-\frac{2N(x,0)}{x}\biggr]
\left({1\over N}{\hat{s}^2+\hat{u}^2\over \hat{s}\hat{u}^2}\right), 
\label{twist3DP}
\eeq
\normalsize
where $f_a(x')$ is the twist-2 unpolarized quark density and
$\delta_a=1(-1)$ for quark (antiquark).  
As shown in (\ref{twist3DP}), the combinations $O(x)\equiv O(x,x)+O(x,0)$ and
$N(x)\equiv N(x,x)-N(x,0)$ appear in the
cross section accompanying the common partonic hard cross section 
which is the same as the twist-2 hard cross section
for the $qg\to q\gamma$ scattering.  
The origin of this simplification 
can be clearly understood in terms of the master formula
for the contribution from the three-gluon correlation functions developed in \cite{KTY11,KY11,KY11direct}. 
We also note that the above result (\ref{twist3DP}) differs from the previous
study in \cite{Ji92}. 

We have performed a numerical calculation for $A_N^\gamma$ 
based on the models used for $p^\uparrow p\to DX$ in Sec. 3.  
For the models 1 and 2, we calculate the asymmetry $A_N^\gamma$
for the two cases:
Case 1; $O(x)=N(x)$ and 
Case 2; $O(x)=-N(x)$. 
We use GJR08 \cite{GJR08} for 
$f_q(x')$ and the models (\ref{model1}) and (\ref{model2}) with $K_G=0.004$ and
$K_G'=0.001$ which are consistent with the RHIC $A_N^D$ data.
We calculate $A_N^\gamma$ at
the RHIC energy at $\sqrt{S}=200$ GeV and the
transverse momentum of the photon $q_T=2$ GeV, 
setting the scale of all the distribution functions at $\mu=q_T$.

\begin{figure}[h]

\scalebox{0.35}{\includegraphics{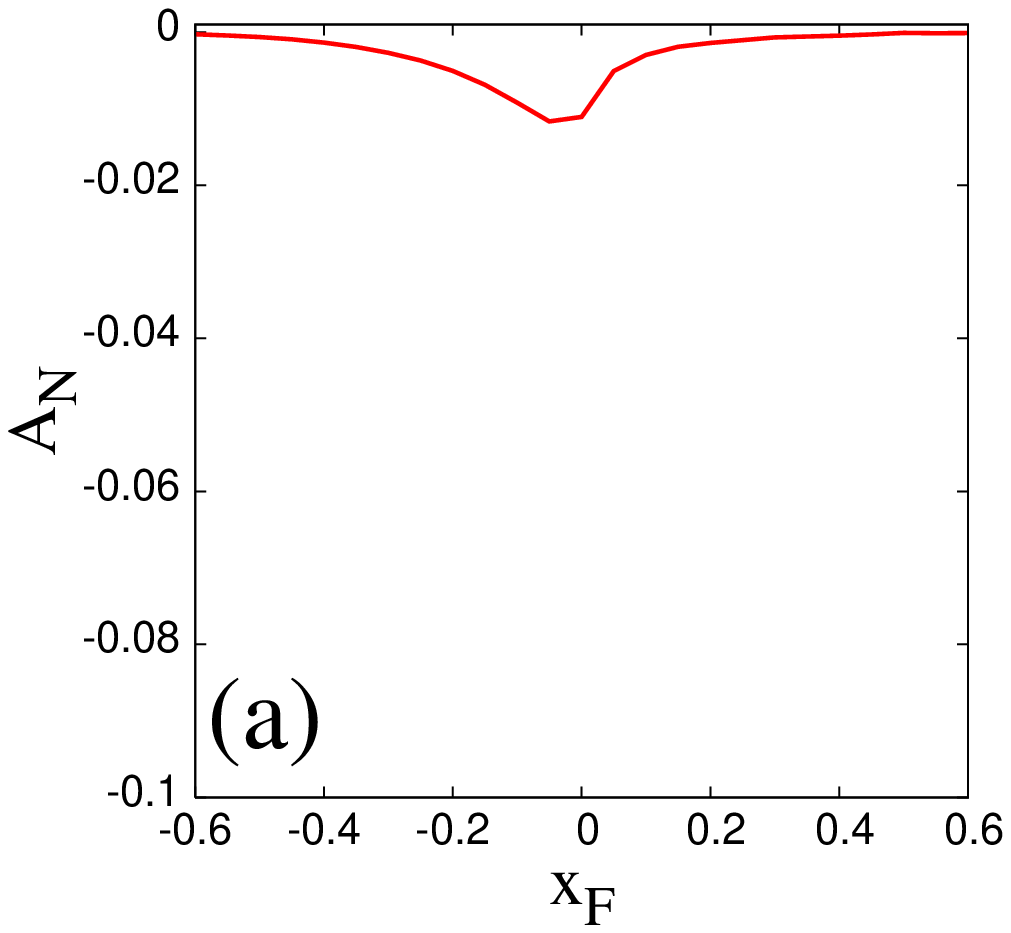}}
\scalebox{0.35}{\includegraphics{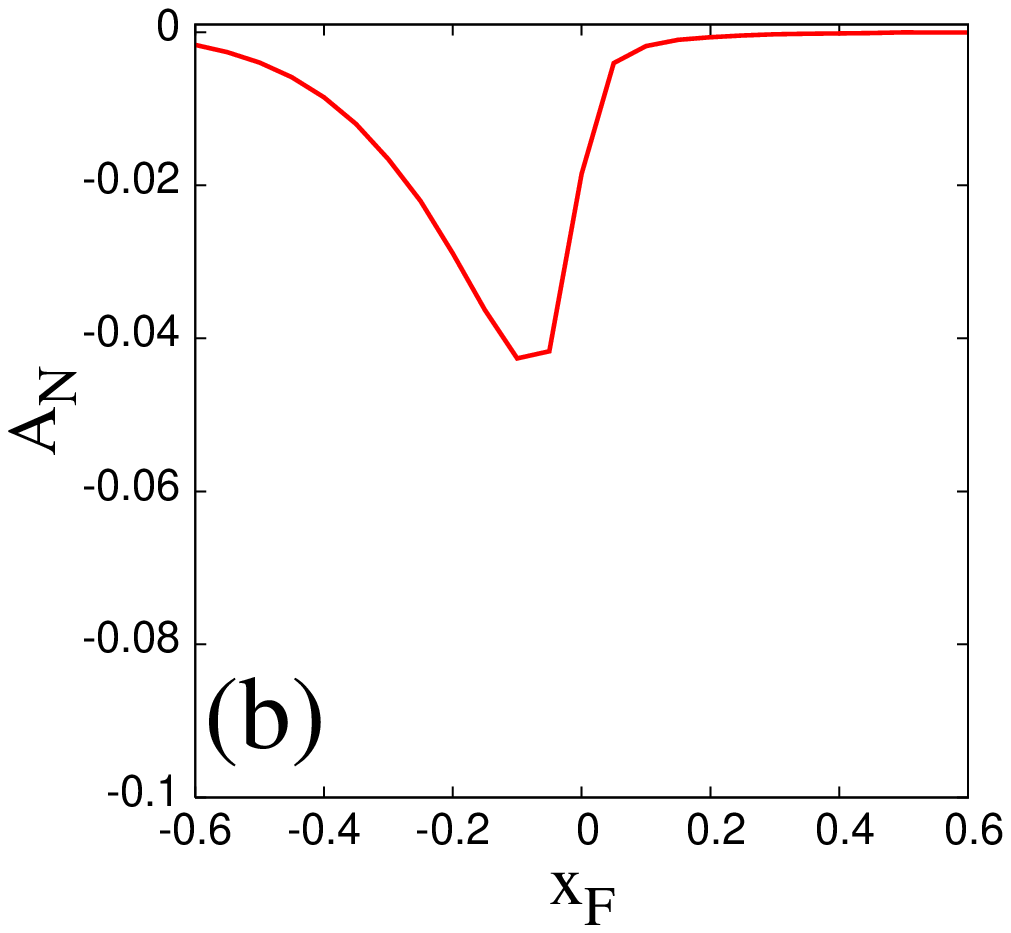}}

\scalebox{0.35}{\includegraphics{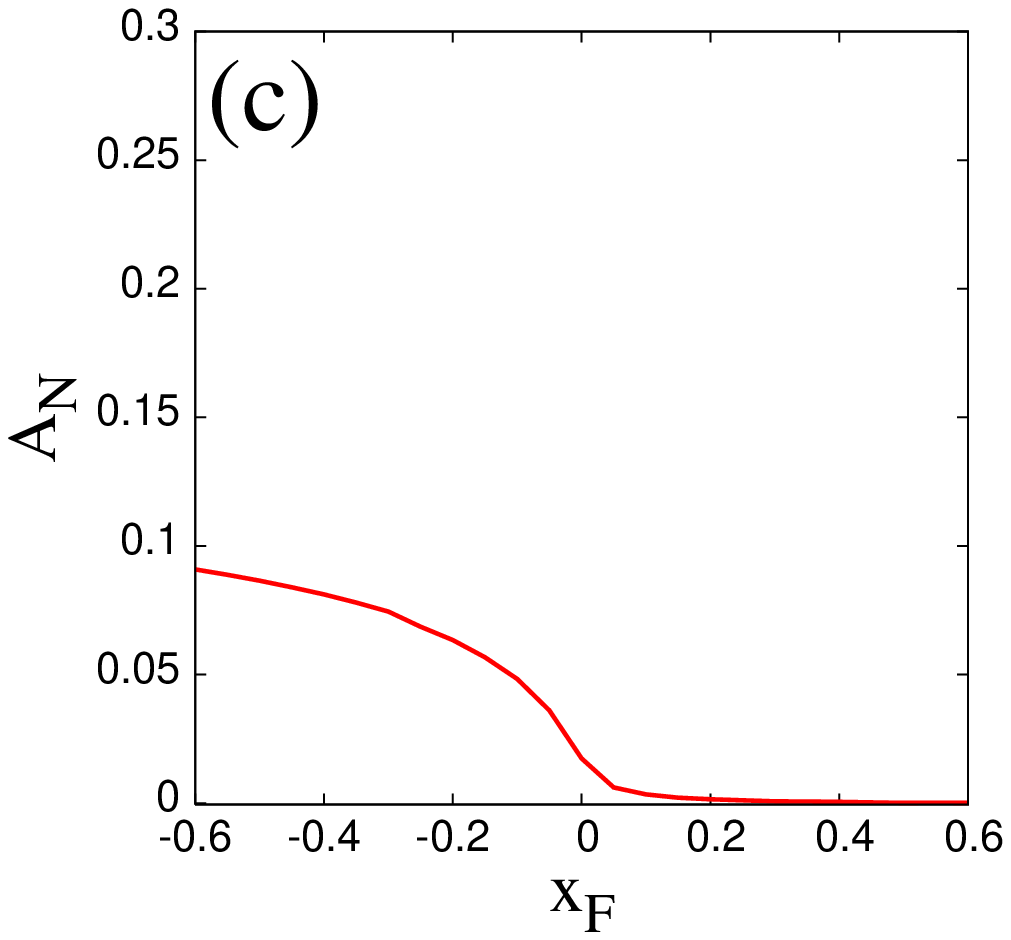}}
\scalebox{0.35}{\includegraphics{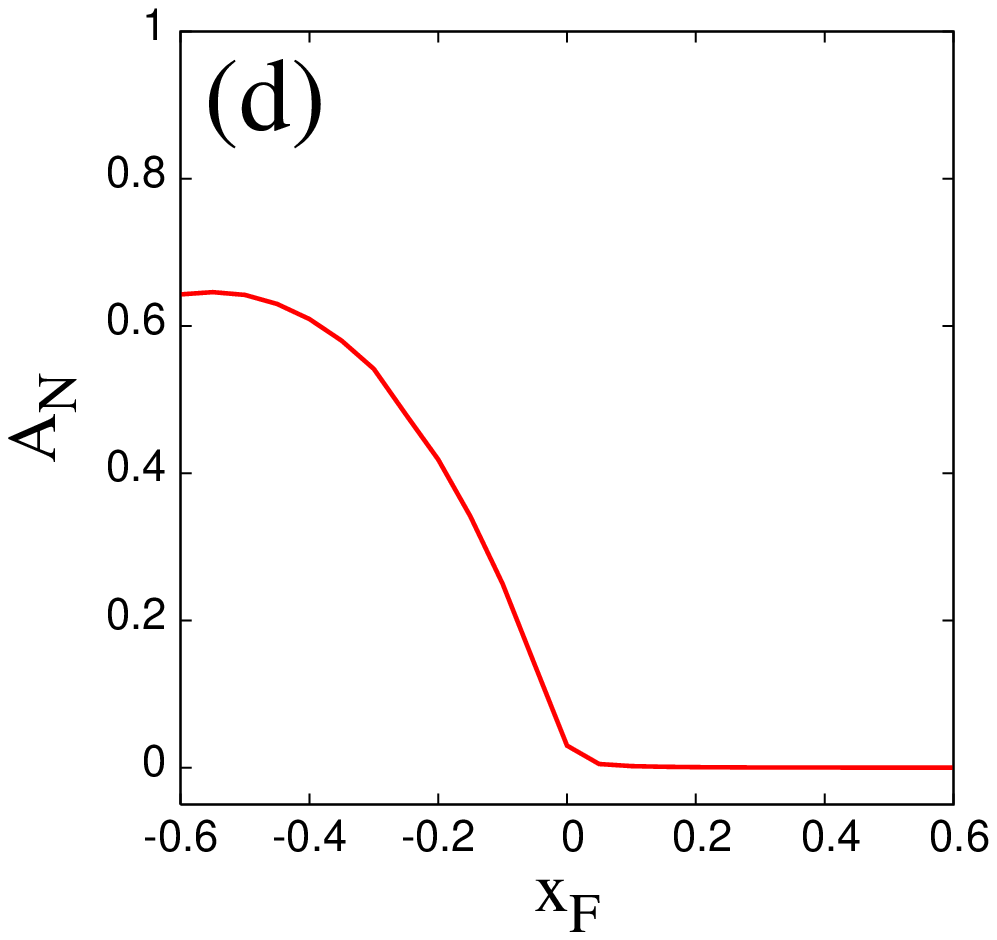}}

 \caption{(a) $A_N$ for Case 1 with Model 1. (b) $A_N$ for Case 1 with Model 2. 
(c) $A_N$ for Case 2 with Model 1. (d) $A_N$ for Case 2 with Model 2. }
 \label{fig:two}

\end{figure}

Fig. 3 shows the result for $A_N^\gamma$ for each case. One can see
$A_N$ at $x_F> 0$ become almost zero regardless of the magnitude of
the three-gluon correlation functions, 
while $A_N^\gamma$
at $x_F<0$ depends strongly on the small-$x$ behavior of the
three-gluon correlation functions as in the case of $p^{\uparrow}p\to DX$.
Even though the derivatives of $O(x)$ and $N(x)$ contribute, 
$A_N$ is tiny at $x_F>0$ due to the small partonic cross section. 
At $x_F<0$, large-$x'$ region of the unpolarized quark distributions
and the small-$x$ region of the three-gluon distributions are relevant.  
For the above case 1, only antiquarks in the unpolarized nucleon are active
and thus lead to small $A_N^\gamma$ as shown in Figs. 3(a) and (b).  
On the other hand, 
for the case 2, quarks in the unpolarized nucleon are active and thus
lead to large $A_N^\gamma$ as shown in  Figs. 3(c) and (d).  
Therefore $A_N^\gamma$ at $x_F<0$ for the direct photon
production could provides us with an important information on the relative sign between $O(x)$
and $N(x)$.

To summarize, we have studied the SSA for $p^\uparrow p\to DX$, $p^\uparrow p\to \gamma^*X$ 
and $p^\uparrow p\to \gamma X$ induced by the three-gluon correlation functions
in the polarized nucleon.  
Combined with the known result for the contribution from the quark-gluon correlations,
this complete the leading-order twist-3 cross sections for these processes.  
We have also presented a model calculation for the asymmetry
for the $p^\uparrow p\to DX$ and $p^\uparrow p\to \gamma X$
at the RHIC energy, showing the sensitivity
of the asymmetry to the form of the three gluon-correlation functions.




\begin{theacknowledgments}
This work is supported by the Grand-in-Aid for Scientific Research
(No. 23540292 and No. 22.6032) from the Japan Society for the Promotion of Science.
\end{theacknowledgments}



\bibliographystyle{aipproc}   

\bibliography{sample}

\IfFileExists{\jobname.bbl}{}
 {\typeout{}
  \typeout{******************************************}
  \typeout{** Please run "bibtex \jobname" to optain}
  \typeout{** the bibliography and then re-run LaTeX}
  \typeout{** twice to fix the references!}
  \typeout{******************************************}
  \typeout{}
 }

\end{document}